\documentstyle[epsf,prl,aps,floats]{revtex}

\begin{document}
\draft
\wideabs{

\title{Zero temperature metal-insulator transition in the \\
infinite-dimensional Hubbard model}

\author{R.~Bulla}
\address{Theoretische Physik III, Institut f\"ur Physik, Universit\"at
Augsburg, 86135 Augsburg, Germany}

\maketitle

\begin{abstract}
The zero temperature transition from a paramagnetic
metal to a paramagnetic insulator is investigated in the Dynamical Mean 
Field Theory for the Hubbard model. The self-energy of the 
effective impurity Anderson model (on which the Hubbard model is
mapped) is calculated using Wilson's Numerical Renormalization Group
method. 
Results for quasiparticle weight, spectral function and
self-energy are discussed for Bethe and hypercubic lattice.
In both cases, the metal-insulator transition is found to occur
via the vanishing of a quasiparticle resonance 
which appears to be isolated from the Hubbard bands.
\end{abstract}


}

The Mott-Hubbard metal-insulator transition \cite{Mott,BUCH}
is one of the most fascinating phenomena of strongly correlated
electron systems. This transition from a paramagnetic metal to
a paramagnetic insulator is found in various transition metal
oxides, such as $\rm V_2O_3$ doped with Cr \cite{McW}.
The mechanism driving the Mott-Hubbard transition is believed to be the
local Coulomb repulsion $U$ between electrons on a same lattice site, although
the details of the transition should also be influenced by lattice degrees
of freedom. 
Therefore,
the simplest model to investigate the correlation driven metal-insulator
transition is the Hubbard model \cite{Hubbard,Gut,Kan}
\begin{equation}
   H = -t\sum_{<ij>\sigma} (c^\dagger_{i\sigma} c_{j\sigma} +
                   c^\dagger_{j\sigma} c_{i\sigma}) +
         U\sum_i c^\dagger_{i\uparrow} c_{i\uparrow}
            c^\dagger_{i\downarrow} c_{i\downarrow}  ,
\label{eq:H}
\end{equation}
where $c^\dagger_{i\sigma}$ ($c_{i\sigma}$) denote creation
(annihilation) operators for a fermion on site $i$, $t$ is the
hopping matrix element and the sum $\sum_{<ij>}$ is restricted
to nearest neighbors.
Despite its simple structure, the solution of this model turns out to
be an extremely difficult many-body problem. The situation is particularly
complicated near the metal-insulator transition where $U$ and the
bandwidth are
roughly of the same order and perturbative schemes (in $U$ or $t$)
are not applicable.

With the recent development of the Dynamical Mean Field Theory
(DMFT) \cite{MVPRLdinfty,PJF,Georges} a very detailed analysis
of the phase diagram of the infinite dimensional Hubbard model
became possible. The Iterative Perturbation Theory (IPT) 
results of \cite{Georges} gave a first order metal-insulator transition
at finite temperatures.
The transition is occuring within
a coexistence region of metallic and
insulating solutions extending from $T\!=\!0$ up to 
$T^\ast\!\approx\! 0.02 W$ ($W$: bandwidth). 
On approaching the metal-insulator transition
from the metallic side (i.e.\ on increasing $U$),
the authors of  \cite{Georges} found a
quasiparticle peak with vanishing spectral weight which becomes isolated
from upper and lower Hubbard bands. A consequence of this result is that the
opening of the gap and the vanishing of the quasiparticle peak do
{\it not} happen at the same critical $U$. The possibility of this scenario
has been questioned by various authors \cite{BUCH,David,Kehrein,NG}.
The criticism is partly based on the fact that the IPT is essentially 
second order perturbation theory in $U$ (although iterated due to
the selfconsistency appearing in the DMFT) whereas the metal-insulator
transition happens at $U$-values of the order of the bandwidth.

Non-perturbative methods are clearly needed to clarify the situation.
At finite temperatures, the Quantum Monte Carlo method
(QMC) should give reliable results and recent QMC
calculations by J.\ Schlipf et al.\ \cite{Sch98} gave 
no indications for a first order transition at finite $T$. 
The experimentally found first order transition in certain transition
metal oxides can therefore not be due to electronic correlations
as modeled in (1) and lattice degrees of freedom will certainly play a role
at the transition.

At zero temperature, the isolation of the quasiparticle peak and the
appearance of a `preformed gap' has been shown by S.\ Kehrein 
\cite{Kehrein} to be in contradiction to a skeleton diagram expansion.
Also, no `preformed gap' has been seen in
calculations based on the Random Dispersion Approximation (RDA) 
\cite{NG} where
the opening of the gap and the vanishing of the quasiparticle peak was 
found to happen at the same critical $U$. 
The results for the gap and the quasiparticle weight are obtained 
in the RDA from
finite size scaling of Exact Diagonalization results of clusters with
up to 14 sites. 

Both QMC and RDA are non-perturbative approaches which can, in principle, be
applied for arbitrary low temperatures (the resolution
of the low-frequency behavior in QMC and RDA 
is limited by the number of time slices or number of sites, respectively).
The only non-perturbative approach
which is presently able to cover the very low temperature regime 
directly in the thermodynamic limit is the 
Numerical Renormalization Group Method (NRG).
This method has been introduced by Wilson for the
Kondo problem \cite{Wil75} and has been
applied by Krishna-murthy et al. to
the impurity Anderson model \cite{Kri80}.
It has been later shown that the NRG allows 
for a very accurate calculation of dynamical properties of various impurity
models
\cite{Sak89,Cos94}. 
This is important because in the DMFT the self-energy (or equivalently
the single-particle spectral function) of an effective
impurity model has to be calculated in the full frequency regime
(for first
applications to the Hubbard model see \cite{Sakai,BHP}).
One therefore expects that the NRG gives equally accurate results
for the effective impurity Anderson model appearing in the DMFT.
However, due to the lack of exact results for, e.g., the metal-insulator
transition in the Hubbard model, this cannot be proven so far.

Here we concentrate on the Mott-Hubbard
metal-insulator transition at zero temperature and half-filling.
At $T\!=\!0$, this transition is usually hidden by the tendency of the model
to form an antiferromagnetic groundstate (as long as no frustration
by, e.g., longer range hopping is included). 
The results are discussed for both
the Bethe lattice with infinite coordination number and the 
infinite dimensional hypercubic lattice.
The hopping matrix element in the hamiltonian 
(\ref{eq:H})  is scaled as
 $t\!=\!t^\ast/\sqrt{\cal Z}$ with $\cal Z$ the number of nearest
neighbors. In the following, we set $t^\ast\!=\!1$ as the unit for the
energy scale. The resulting free densities of states for
Bethe and hypercubic lattice are
\begin{eqnarray}
   \rho_{\rm B}(\varepsilon)&=&\frac{1}{2\pi}\sqrt{4-\varepsilon^2}
        \ \ \ : |\varepsilon| \le 2 \label{eq:semi} , \\
   \rho_{\rm hc}(\varepsilon)&=&\frac{1}{\sqrt{2\pi}} \exp\left(
             -\frac{\varepsilon^2}{2}\right)\label{eq:gauss} .
\end{eqnarray}
The effective bandwidth $W\!=\!4\sqrt{ \int {\rm d}\varepsilon
      \rho(\varepsilon)   \varepsilon^2} $ is $W\!=\!4$ for both
$\rho_{\rm B}$ and $\rho_{\rm hc}$ (the factor $4$ is chosen
so that the $W$ corresponds to the actual bandwidth of the semi-elliptic
density of states  $\rho_{\rm B}$).

\begin{figure}[tp]
\epsfxsize=3.0in
\epsffile{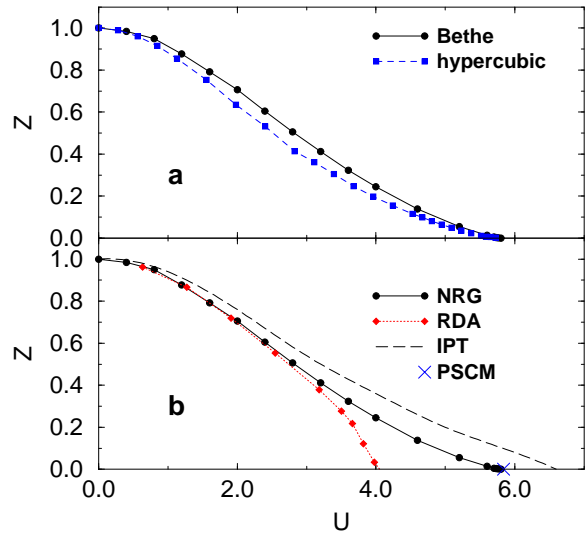}
\caption[]{$U$ dependence of the quasiparticle weight $Z$; a:
  comparison of the NRG results for Bethe and hypercubic lattice and
  b: comparison of the NRG results with
     results from the RDA and the IPT (for the Bethe lattice). Also
     shown is the result for the critical $U$ from the PSCM.
    }
\label{fig:Z}
\end{figure}

Fig.\ \ref{fig:Z}a shows the 
$U$-dependence of the quasiparticle weight
\begin{equation}
Z=\frac{1}{1-\frac{\partial \Re e\Sigma(\omega)}{\partial
\omega}|_{\omega=0}},
\end{equation}
for both lattices.
Despite the different lattice structure,
the critical value of $U$ is 
approximately the same
for both Bethe and hypercubic lattice, $U_{\rm c,B}\approx 5.88 \!=\! 1.47W$
and $U_{\rm c,hc}\approx 5.80 \!=\! 1.45W$.
The different behavior of the $Z(U)$-curves for small values of $U$
can be understood from second order perturbation theory which gives
$Z(U)\!=\! 1.0 - 0.082\ U^2 + {\cal O}(U^4)$ for the
Bethe lattice \cite{NG} and
$Z(U)\!=\! 1.0 - 0.12\ U^2 + {\cal O}(U^4)$ for the hypercubic lattice.

Fig.\ \ref{fig:Z}b shows the
NRG-result for the Bethe lattice together with results 
from calculations using the Random Dispersion Approximation
(RDA) \cite{NG} and the IPT \cite{Georges}. The NRG and RDA
results agree very well up to $U\!\approx\!2.5$ but the NRG gives
a long tail in the $Z(U)$-curve ending at a critical 
$U_{\rm c}$ which is considerably larger than the 
$U_{\rm c,RDA}\approx W$. 
In my view, the difference for 
$U\!>\!2.5$ may be a consequence of the small system sizes
presently taken into account in the RDA. 

The $Z(U)$-curve from the IPT starts to deviate from the NRG
result already for very small values of $U$.
It has been found earlier
that the critical $U$ is overestimated by the IPT 
($U_{\rm c,IPT}\approx  1.65W$ \cite{Georges}) and that the
Projective Self-consistent Method (PSCM) \cite{Georges,Moeller} gives
a lower $U_{\rm c,PS}\approx  1.46W$ which is in remarkable agreement 
with the NRG-result.
The $Z(U)$-curve from the QMC for small finite
temperatures and a Bethe lattice density of states
(not shown here) agrees well with the NRG result
for $T\!=\!0$ up to $U\!\approx\!4.5$ \cite{Sch98}.
The critical values obtained from the QMC 
(e.g.\  $U_{\rm c,QMC}= (1.26\pm0.01)W$ for $T\!=\!1/30$)
are smaller than those from the NRG 
and the $U_{\rm c,QMC}(T)$-curve shows a negative
slope, a consequence of the higher spin-entropy of the insulating 
phase.

\begin{figure}[tp]
\epsfxsize=3.0in
\epsffile{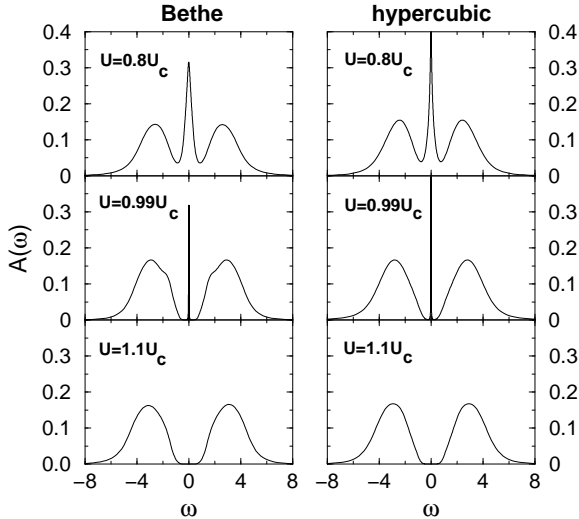}
\caption[]{
  Spectral functions for Bethe and hypercubic lattice for various values
  of $U$. In both cases, a narrow quasiparticle peak develops at the
  Fermi level which vanishes at the critical $U_{\rm c}$.
  }
\label{fig:spectra}
\end{figure}

The spectral functions 
$A(\omega)$ for Bethe and hypercubic lattice are compared
in Fig.\ \ref{fig:spectra} for $U\!=\!0.8U_{\rm c}$, $U\!=\!0.99U_{\rm c}$
and $U\!=\!1.1U_{\rm c}$ (for details of the numerical calculations,
see \cite{BHP}). 
Although the semi-elliptic density of states $\rho_{\rm B}$ is confined to
the interval $[-2,2]$, whereas the gaussian density of states 
$\rho_{\rm hc}$  has no cutoff, the structures appearing in the spectral
functions are very similar. In the metallic phase (for large enough values
of $U$) the spectral function shows the typical three-peak structure
with upper and lower Hubbard bands centered at $\pm U/2$
and a quasiparticle
peak at the Fermi level. For $U\!=\!0.99U_{\rm c}$, the quasiparticle peak
in both Bethe and hypercubic 
lattice seems to be isolated (within the numerical accuracy)
from the upper and lower Hubbard
bands, similar to what has been observed in the IPT calculations
for the Bethe lattice
\cite{Georges}. Consequently, the gap appears to open discontinuously
at the critical $U$.
Note that, due to the broadening of the spectra \cite{BHP}, an accurate
resolution of the high energy features, e.g., the band edges of the Hubbard
bands, is not possible.

The term `preformed gap' is frequently used to describe 
the behavior seen in Fig.\ \ref{fig:spectra}
 although it is not clear whether a `preformed gap'
only
means a strong suppression of spectral weight between the Hubbard bands
(as seen in both NRG and IPT) or an exact vanishing of the spectral
function in a finite interval. Using the latter definition,
it is not possible to decide within a numerical approach 
(like NRG, QMC or IPT) whether
the system shows a `preformed gap' or not. 


\begin{figure}[tp]
\epsfxsize=3.2in
\epsffile{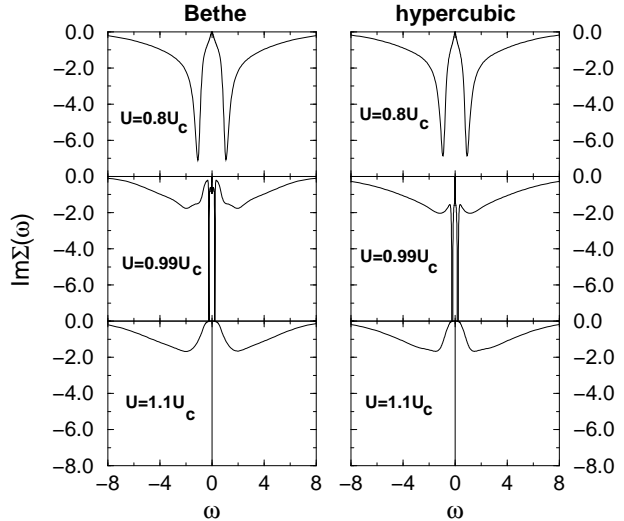}
\caption[]{
  Imaginary part of the self-energy for Bethe and hypercubic lattice
  for the same parameters as in Fig.\ \ref{fig:spectra}.
  }
\label{fig:ImS}
\end{figure}

The three-peak structure in the spectral function and the isolation
of the quasiparticle peak near the transition have important consequences
for the behavior of the self-energy. Fig.\ \ref{fig:ImS} shows
the imaginary part of the self-energy for the same parameters as
in Fig.\ \ref{fig:spectra}. In the
insulating regime ($U\!=\!1.1U_{\rm c}$) the self-energy has a 
pole at zero frequency $\Sigma(z)\!=\!\frac{\alpha}{z}+
\Sigma_{\rm rem}(z)$ ($z\!=\!\omega+i0^{+}$, $\Sigma_{\rm rem}$ denotes
the remaining part of the self-energy).
There are several possibilities how the $1/z$-term 
in $\Sigma(z)$ develops when the transition
is approached from the metallic side.
One possibility would be that $\alpha$ is zero at the transition
($U\!=\!U_{\rm c}$)
and then increases continuously with increasing $U$.
This imposes some constraints on the form of the spectral function
because the weight $\alpha$ of the pole in the insulating regime is given by
\begin{equation}
  \alpha^{-1} = \int_{-\infty}^{\infty} {\rm d}\omega 
  \frac{A(\omega)}{\omega^2}
.
\end{equation} 
If one assumes a powerlaw $A(\omega)\propto|\omega |^r$ for small
$\omega$, the exponent has to satisfy $r\le 1$ for $\alpha$ to vanish
at the transition.

The other possibility is what is seen in the
NRG-results for the Bethe and hypercubic lattice in
Fig.\ \ref{fig:ImS}.
The $1/z$-term emerges from a two-peak structure in the imaginary
part of the self-energy. 
The weight of the peaks is roughly independent
of $U$, while the position 
and the width  vanish with $U\to U_{\rm c}$ (the position is proportional
to $\sqrt{Z}$). 
At the transition, the two peaks
collapse  and give rise to a single pole with weight $\alpha$
(these features have already been discussed in \cite{Kehrein,Zhang};
note that the two-peak structure in Im$\Sigma$ does not imply the existence
of poles in $\Sigma(z)$).
This behavior is common to both lattice types studied
here and it is only the $U$-dependence of the width $x$, that
differs between Bethe and hypercubic lattice. 

Note that the vanishing of the quasiparticle peak in the standard
single impurity Anderson model (which occurs for 
$U/\pi\Delta\to\infty$ \cite{Hewson})
is also associated with the collapse of a two-peak structure in the 
self-energy. This is observed both in the wide-band and the narrow-band
limit (see \cite{Walter} for a discussion of the latter case).

The two-peak structure is related to the typical three-peak structure
(quasiparticle peak plus upper and lower Hubbard bands)
in the spectral functions for both single impurity Anderson model
 and the infinite
dimensional Hubbard model.
In both models, one has 
the relation 
\begin{equation}
\Sigma(z) = z-\varepsilon_{\rm d}
-\Delta(z) - \frac{1}{G(z)},
\end{equation}
 with the hybridization function $\Delta(z)$.
The self-energy develops peaks at the frequencies where 
real- and imaginary part of $G(z)$ are small, which is the region between
the quasiparticle peak and the Hubbard bands. 
Therefore, all calculations for the infinite
dimensional Hubbard model which give a well-pronounced three-peak structure
in $A(\omega)$ necessarily produce the two-peak structure in the self-energy
(examples are calculations from the QMC \cite{PJF}
and the Non Crossing
Approximation \cite{Pru93}; at finite temperatures, the two-peak structure
is broadened).

We now turn to an additional feature seen in both IPT and NRG
calculations: the coexistence of metallic and insulating solutions
in an interval $U_{\rm c,1}<U<U_{\rm c,2}$.
Starting from $U\!=\!0$, the metal
to insulator transition occurs at the critical $U_{\rm c,2}$
with the vanishing of the quasiparticle peak. Starting 
from the insulating side, the insulator
to metal transition happens at $U_{\rm c,1}<U_{\rm c,2}$
(the NRG and IPT give 
$U_{\rm c,1}\approx 1.25W$ for the Bethe lattice and the NRG gives
$U_{\rm c,1}\approx 1.15W$ for the hypercubic lattice).

The coexistence of metallic and insulating solutions 
is probably connected to  
the structure of the self-energy at $T\!=\!0$. When $U$ is reduced
below $U_{\rm c,2}$, the $\delta$-function peak in the imaginary
part of the self-energy 
does not split into the two-peak structure which is found for the metallic
solution. 
The $\delta$-function peak only vanishes when its weight $\pi\alpha$
vanishes which happens at a lower value of $U$ (the $U_{\rm c,1}$).
The $U$-dependence of $\alpha$ near $U_{\rm c,1}$ is difficult to
determine and it is presently not clear whether $\alpha$ vanishes
continuously or not. 

The physical solution of the DMFT equations in the coexistence region is the
one with the lower energy which turns out to be the metallic one,
in agreement with \cite{Georges} (near the transition, the energy difference
becomes too small to decide which solution has the lower
energy, so that a small uncertainty remains near $U_{\rm c,2}$).
The coexistence of solutions therefore does not play a role
at $T\!=\!0$  and can be neglected.

In conclusion, we have investigated the zero-temperature metal-insulator
transition in the Hubbard model for both Bethe and hypercubic lattice
using a non-perturbative approach, the Numerical Renormalization Group 
method.
The NRG calculations show that the details of the transition are very
similar in both cases despite the different lattice structure.
In the Bethe lattice case, the result for the critical $U$ 
is in remarkable agreement with the result from the 
Projective Self-consistent Method.
The NRG-results (in particular, the two-peak structure in the imaginary
part of the self-energy near the critical $U$)
cannot be explained within a skeleton diagram expansion
as shown in \cite{Kehrein}. Potential problems due
to the fact that the derivation of the DMFT is based on
such a skeleton diagram expansion\cite{Walter} have still to be clarified.

In order to bridge the gap between $T\!=\!0$ and the lowest 
temperatures accessible to the QMC-method, the NRG has to be
extended to finite temperatures (work on this is in progress).
This will allow to study the metal-insulator transition in the whole
temperature range using non-perturbative methods.

The author would like to thank P.\ van Dongen,
F.\ Gebhard, 
A.\ C.\ Hewson, V.\ Janis, S.\ Kehrein, D.\ Logan,
R.\ Noack, R.\ Pietig, Th.\ Pruschke and D.\ Vollhardt
for many helpful discussions.


\begin{references}

\bibitem{Mott} N.~F.~Mott, Proc.~Phys.~Soc.~London~A~{\bf 62}, 
416 (1949); {\sl Metal-Insulator Transitions}, 2nd ed.\ 
(Taylor and Francis, London, 1990).

\bibitem{BUCH} F.~Gebhard, {\sl The Mott Metal-Insulator Transition}, 
Springer Tracts in Modern Physics Vol.~137 (Springer, Berlin, 1997).

\bibitem{McW} D.\ B.\ McWhan and J.\ P.\ Remeika,
  Phys.~Rev.\ B {\bf 2}, 3734 (1970);
D.\ B.\ McWhan, A.\ Menth, J.\ P.\ Remeika,
   Q.\ F.\ Brinkman and T.\ M.\ Rice,
  Phys.~Rev.\ B {\bf 7}, 1920 (1973).

\bibitem{Hubbard} J.~Hubbard, Proc.~R.~Soc.~London A~{\bf 276}, 238 (1963).

\bibitem{Gut} M.\ C.\ Gutzwiller, Phys.~Rev.~Lett.~{\bf 10}, 59 (1963).

\bibitem{Kan} J.\ Kanamori, Prog.\ Theor.\ Phys.\ {\bf 30}, 275 (1963).

\bibitem{MVPRLdinfty} W.~Metzner and D.~Vollhardt, 
Phys.~Rev.~Lett.~{\bf 62}, 324 (1989); for an introduction,
see D.~Vollhardt, Int.\ J.~Mod.\ Phys.~B~{\bf 3}, 2189 (1989).

\bibitem{PJF} M.~Jarrell, Phys.~Rev.~Lett.~{\bf 69}, 168 (1992);
T.~Pruschke, M.~Jarrell, and J.~K.~Freericks,
Adv.~Phys.~{\bf 44}, 187 (1995).

\bibitem{Georges}  A.~Georges, G.~Kotliar, W.~Krauth, and
M.~J.~Rozenberg, Rev.\ Mod.\ Phys.~{\bf 68}, 13 (1996).

\bibitem{David} D.~E.~Logan and P.~Nozi\`eres,
Phil.~Trans.~R.~Soc.\ London A~{\bf 356}, 249 (1998).
 
\bibitem{Kehrein} S.~Kehrein, 
  Phys.~Rev.~Lett.~{\bf 81}, 3912 (1998).

\bibitem{NG} R.\ Noack and F.\ Gebhard, preprint (cond-mat/9810222).

\bibitem{Sch98} J.\ Schlipf, M.\ Jarrell, P.\ G.\ J.\ van Dongen,
    S.\ Kehrein, N.\ Bl\"umer, Th.\ Pruschke and D.\ Vollhardt,
    preprint (1999) and private communication.

\bibitem{Wil75}
  K.~G.~Wilson,
  Rev.~Mod.~Phys.~{\bf 47}, 773 (1975).

\bibitem{Kri80}
  H.~R.~Krishna-murthy, J.~W.~Wilkins and K.~G.~Wilson,
  Phys.\ Rev.\ B {\bf 21}, 1003 \& 1044 (1980).

\bibitem{Sak89}
  O.\ Sakai, Y.\ Shimizu and T.\ Kasuya, 
  J.\ Phys.\ Soc.\ Jpn.\ {\bf 58}, 3666 (1989).

\bibitem{Cos94}
  T.\ A.\ Costi, A.\ C.\ Hewson and V.\ Zlati\'c, 
  J.\ Phys.: Cond.\ Matter {\bf 6}, 2519 (1994).

\bibitem{Sakai}
  O.\ Sakai and Y.\ Kuramoto, 
  Sol.\ Stat.\ Comm.\ {\bf 89}, 307 (1994).


\bibitem{BHP}  R.\ Bulla, A.\ C.\ Hewson and Th.\ Pruschke,  
  J.\ Phys.: Cond.\ Matter {\bf 10}, 8365 (1998).


\bibitem{Moeller} G.\ Moeller, Q.\ Si, G.\ Kotliar, M.\ Rozenberg  and
  D.\ S.\ Fisher,
  Phys.\ Rev.\ Lett.\ {\bf 74}, 2082 (1995).


\bibitem{Zhang} X.\ Y.\ Zhang, M.\ J.\ Rozenberg and G.\ Kotliar
      Phys.\ Rev.\ Lett.\ {\bf 70}, 1666 (1993).

\bibitem{Hewson} 
  A.\ C.\ Hewson 
  {\it The Kondo Problem to Heavy Fermions} 
  (Cambridge: Cambridge Univ. Press, 1993).

\bibitem{Walter} W.\ Hofstetter, S.\ Kehrein, preprint 
     cond-mat/9812427 (1998).

\bibitem{Pru93}
  Th.\ Pruschke, D.\ L.\ Cox, M.\ Jarrell,
  Phys.\ Rev.\ B {\bf 47}, 3553 (1993).

\end{references}
\end{document}